\documentclass{appolb}

\usepackage{calc}
\usepackage{graphicx}
\usepackage{dcolumn}
\usepackage{bm}
\usepackage{slashed}
\usepackage{amsmath,graphicx}
\usepackage[colorlinks=true,linktocpage=true,linkcolor=blue,citecolor=blue]{hyperref}
\usepackage{float}
\usepackage{nicefrac}
\usepackage[normalem]{ulem}
\usepackage{amsmath}
\usepackage{subfigure}
\usepackage{float} 
\usepackage{bbold}
\usepackage{cite} 

\def\be{\begin{equation}}
	\def\ee{\end{equation}}
\newcommand{\bel}[1]{\begin{eqnarray}\label{#1}}
	\newcommand{\eel}{\end{eqnarray}}
\def\barr{\begin{array}}
	\def\earr{\end{array}}
\def\beq{\begin{eqnarray}}
	\def\eeq{\end{eqnarray}}
\def\bfig{\begin{figure}}
	\def\efig{\end{figure}}

\newcommand{\rf}[1]{Eq.~(\ref{#1})}

\newcommand{\rfn}[1]{(\ref{#1})}

\def\a{\alpha}
\def\b{\beta}
\def\g{\gamma}
\def\d{\delta}

\def\HP{\hphantom{\alpha}} 

\newcommand{\sh}[1]{\sinh#1}
\newcommand{\ch}[1]{\cosh#1}

\newcommand{\olra}{\overleftrightarrow}

\def\cA{{\cal A}}
\def\cB{{\cal B}}
\def\cC{{\cal C}}

\def\cN{{\cal N}}
\def\cE{{\cal E}}
\def\cP{{\cal P}}

\def\cNN{{\cal N}_{(0)}}
\def\cEN{{\cal E}_{(0)}}
\def\cPN{{\cal P}_{(0)}}

\def\be{\begin{equation}}
\def\ee{\end{equation}}
\def\ba{\begin{eqnarray}}
\def\ea{\end{eqnarray}}   

\def\a{\alpha}
\def\b{\beta}
\def\g{\gamma}
\def\d{\delta}

\def\n0{n_{(0)}}
\def\e0{\varepsilon_{(0)}}
\def\P0{P_{(0)}}

\begin{document} 
\title{CONFORMAL TRANSFORMATIONS OF CONSERVATION EQUATIONS IN SPIN HYDRODYNAMICS
\thanks{Presented at the XXVII Cracow EPIPHANY Conference on Future of particle physics, Krakow, Poland, January 07-10, 2021.} 
}
\author{Rajeev Singh
\address{Institute of Nuclear Physics Polish Academy of Sciences, 31-342 Krak\'ow}
}
\maketitle
\begin{abstract} 
We study and analyse the conformal transformations of different conservation laws in the spin hydrodynamics framework.
\end{abstract}
\section{Introduction}
Relativistic fluid dynamics has been proved to be a successful theory for ultra-relativistic heavy-ion collisions~\cite{Florkowski:2010zz,Gale:2013da,Romatschke:2017ejr,Alqahtani:2017mhy}. Spin polarization measurements of $\Lambda$ hyperons done recently indicated that the we should include space-time evolution of spin in the framework of relativistic fluid dynamics~\cite{STAR:2017ckg, Adam:2018ivw,Adam:2019srw,Niida:2018hfw,Acharya:2019vpe,Acharya:2019ryw}.
Formulation of such a spin fluid dynamics formalism was studied first in Ref.~\cite{Florkowski:2017ruc}, which gave rise to many studies~\cite{Florkowski:2017dyn,Florkowski:2018myy,Becattini:2018duy,Florkowski:2018ahw,Florkowski:2019voj,Florkowski:2019qdp,Bhadury:2020puc,Bhadury:2020cop,Florkowski:2021pkp,Florkowski:2018fap,Tinti:2020gyh}. Contrary to theoretical studies done on spin which considered spin-vorticity coupling to be the reason for the spin polarization at freeze-out~\cite{Becattini:2013fla,Becattini:2016gvu,Karpenko:2016jyx,Xie:2017upb,Li:2017slc,Ivanov:2019wzg,Becattini:2019ntv,Zhang:2019xya,Fukushima:2020ucl,Gao:2020lxh,Gao:2020pfu,Montenegro:2017rbu,Prokhorov:2018bql,Yang:2018lew,Li:2019qkf,Liu:2019krs,Ambrus:2019ayb,Sheng:2019kmk,Hattori:2019lfp,Prokhorov:2019yft,Deng:2020ygd,Yang:2020hri,Gallegos:2020otk,Shi:2020htn,Garbiso:2020puw,Gallegos:2021bzp,Weickgenannt:2021cuo,Sheng:2021kfc,Wu:2019eyi,Becattini:2007sr,Florkowski:2019gio,Xie:2019jun,Liu:2020dxg,Becattini:2020sww,Liu:2020flb,Gao:2020vbh,Weickgenannt:2020aaf,Fukushima:2020qta,Becattini:2020ngo,Speranza:2020ilk,Liu:2021uhn,Fu:2021pok,Becattini:2021iol,Becattini:2007nd,Becattini:2009wh}, 
the spin hydrodynamic formalism has been developed on the basis on conservation laws and local thermodynamic equilibrium, introducing new dynamic quantity namely spin polarization tensor. In this work, we use the Gubser symmetry~\cite{Gubser:2010ze,Gubser:2010ui,Shokri:2018qcu} arguments to find the conformal transformations and criteria for conformal invariance of the conservation laws which are used in the spin hydrodynamics framework~\cite{Florkowski:2018fap,Florkowski:2019qdp,Singh:2020rht}. Throughout the article we use natural units, i.e., $c = \hbar = k_B=1$.
\section{Conservation laws}
Baryon number conservation law is given by $d_\a N^\a(x)  = 0$, where for perfect-fluid case, net baryon current $N^\a$ is defined as
\bel{eq:N}
N^\a = \cN U^\a = 4  \sh(\frac{\mu}{T}) \,\, \cNN~U^\a\, ,
\eel
with $\cN$ and $U^\a$ being the net baryon density and fluid flow vector, respectively.
$\mu$, $T$, and $\cNN$ is baryon chemical potential, temperature, and number density for ideal relativistic gas of classical massive particles~\cite{Florkowski:2010zz}, respectively.

For perfect-fluid dynamics case, energy and linear momentum conservation law is expressed as
\bel{eq:Tcon}
d_\a T^{\a\b}(x) \equiv d_\a [(\cE+\cP) \, U^\a U^\b + \cP g^{\a\b}] = 0\, 
\eel
with the energy density, $\cE=4 \ch(\frac{\mu}{T})\cEN$ and pressure, $\cP=4 \ch (\frac{\mu}{T})\cPN$,
with $\cEN$ and $\cPN$ being the energy density and pressure for ideal relativistic gas of classical massive particles~\cite{Florkowski:2010zz}.

Total angular momentum consists of orbital angular momentum $L^{\a,\b\g}$ and spin angular momentum $S^{\a,\b\g}$ as
\beq
J^{\a,\b\g}&=&L^{\a,\b\g} +S^{\a,\b\g}=x^{\b} T^{\a\g} - x^{\g} T^{\a\b} + S^{\a,\b\g}\,.
\label{eq:L}
\eeq
where the total angular momentum conservation law is $d_\a J^{\a,\b\g}=d_\a S^{\a,\b\g}+ 2 T^{[\b\g]}= 0$.
Symmetric $T^{\mu\nu}$ implies the conservation of spin separately, where the spin tensor is given as~\cite{Florkowski:2019qdp,DeGroot:1980dk}
\beq
S_{\rm GLW}^{\a,\b\g}
=\cC U^\a \omega^{\b\g}+\cA U^\a U^{[\b} k^{\g]} +\cB \big[U^{[\b} \Delta^{\a\d} \omega^{\g]}_{\HP\d}+ U^\a \Delta^{\d[\b} \omega^{\g]}_{\HP\d}
+\Delta^{\a[\b} k^{\g]}\big]
\label{eq:S}
\eeq
with $\omega^{\a\b}(x)$ is spin polarization tensor and $\cA$, $\cB$ and $\cC$ are defined as 
$\cA = 2 {\cal C} -3 {\cal B}$, $\cB =-\frac{\cE+\cP}{2 T z^2 }$, and $\cC= \frac{\cP}{4T}$.
\section{Gubser symmetry and conformal weights}
\label{sec:gubser}
Gubser symmetry~\cite{Gubser:2010ze,Gubser:2010ui} consist of two special conformal
transformations, i.e., ($SO(3)_q$), boosts in the direction of $\eta$, i.e., ($SO(1,1)$) and reflection in the $r-\phi$ plane, i.e., ($Z_2$).
For a system to conserve conformal symmetry, it needs to be invariant under Weyl rescaling \cite{Baier:2007ix,Bhattacharyya:2007vs,Loganayagam:2008is,Gubser:2010ze,Gubser:2010ui} which implies that tensors of type $(m,n)$ transform homogeneously as
\ba
 {\cal{X}}^{\mu_1 ...\mu_m}_{\nu_1 ...\nu_n}(x)\,\rightarrow\,\Omega^{\Delta_{\cal{X}}}{\cal{X}}^{\mu_1 ...\mu_m}_{\nu_1 ...\nu_n}(x)\rightarrow\,e^{-\varphi (x)\Delta_{\cal{X}}}{\cal{X}}^{\mu_1 ...\mu_m}_{\nu_1 ...\nu_n}(x)\, ,
\label{eq:weyl-rescaling}
\ea
where $\Delta_{\cal{X}} = [{\cal{X}}]+m-n$ is the conformal weight of the quantity $\cal{X}$
with $[\cal{X}]$ being its mass dimension, and $m$ and $n$ represents the contravariant and covariant indices, respectively.
For example, the rank 2 dimensionless metric tensor $g_{\mu\nu}$ transforms under Weyl rescaling as \cite{Baier:2007ix,Gubser:2010ui}
\ba 
g_{\mu\nu} \rightarrow \Omega^{-2}\,g_{\mu\nu}\, .
\label{eq:g-weyl}
\ea
Using Eq.~(\ref{eq:g-weyl}) and the normalization of flow vector, one can find $\Delta_{U^\mu}=1$.
Energy density and pressure have mass dimension $[\cE]\equiv[\cP]=4$, hence one can obtain their conformal weight to be $\Delta_{\cE}=\Delta_{\cP}=4$. Similarly, net baryon density has conformal weight $\Delta_{\cN}=3$, and temperature and baryon chemical potential have the same conformal weight $\Delta_T=\Delta_{\mu}=1$, as both have same mass dimension, which is 1.
Conformal weight of energy-momentum tensor can be known using the conformal weight of $\cE$, $\cP$, and $U^\mu$, which results to $\Delta_{T^{\a\b}}=6$, since $\Delta_{x^\b}=0$. One can see from \rf{eq:L} that spin tensor should have same conformal weight as $T^{\mu\nu}$, so, canonical spin tensor~\cite{Weinberg:1995mt}, GLW spin tensor~\cite{DeGroot:1980dk} in \rf{eq:S}, and HW spin tensor~\cite{Speranza:2020ilk,Weickgenannt:2020aaf}, expressed respectively below, should have same conformal weights
 \ba
 \label{eq:scan}
 S_{\rm C}^{\a,\b\g} &=& \frac{i}{8}\bar{\psi}\{\gamma^\a,[\gamma^\b,\gamma^\g]\}\psi\, ,\\
 \label{eq:GLW}
 S^{\alpha,\beta\gamma}_{\rm GLW}&=&\frac{i}{4m}\left(\bar{\psi}\sigma^{\beta\gamma}\olra{\partial}^\alpha\psi-\partial_\rho \epsilon^{\beta\gamma\alpha\rho}\bar{\psi} \gamma^5\psi \right),\\
 \label{eq:HW}
 S_{\rm HW}^{\a,\b\g} &=& S_{\rm C}^{\a,\b\g} -\frac{1}{4m}\left( \bar{\psi} \sigma^{\b\g} \sigma^{\a\rho}\partial_\rho \psi + \partial_\rho \bar{\psi} \sigma^{\a\rho} \sigma^{\b\g}\psi\right),
\ea
with
$\sigma^{\mu \nu}\!\!=\!\!\frac{i}{2}\left[\gamma^{\mu}, \gamma^{\nu}\right]
$, where, spinor $\psi$ and dual spinor $\bar{\psi}\equiv\psi^{\dagger} \gamma_{0}$ have conformal weight $\Delta_\psi = \Delta_{\bar{\psi}} = \frac{3}{2}$ and gamma matrix have conformal weight $\Delta_{\gamma^\mu}=1$~\cite{Kastrup:2008jn,Fabbri:2011ha}, hence $\Delta_{S_{\rm GLW}^{\a\b\g}}=\Delta_{S_{\rm C}^{\a\b\g}}=\Delta_{S_{\rm HW}^{\a\b\g}}=6$.
In similar way, one can obtain conformal weight of net baryon number current, $\Delta_{N^\a}=4$, using conformal weight of $\cN$ and \rf{eq:N}.
We summarize below the transformation rules under Weyl rescaling (for 4D spacetime) using \rf{eq:weyl-rescaling}
\beq
\label{eq:ConfN}
N^\a \rightarrow \Omega^4 ~N^\a, \quad T^{\a\b} \rightarrow \Omega^6 ~T^{\a\b}, \quad S^{\a\b\g} \rightarrow \Omega^6 ~S^{\a\b\g}.
\eeq
\section{Conformal invariance of laws of conservation}
\label{sec:invariance}
%
For 4D conformal fluid dynamics, we intend to find the conformal transformations of the laws of conservation for net baryon number, energy and linear momentum, and spin, which are given below as
\ba
\label{eq:dN}
d_\a N^\a(x) &=& \partial_\a N^\a + \Gamma^\a_{\a\b} N^\b = 0\, , \\
\label{eq:dT}
d_\a T^{\a\b}(x) &=& \partial_\a T^{\a\b} + \Gamma^\a_{\a\lambda} T^{\lambda\b} + \Gamma^\b_{\a\lambda} T^{\a\lambda} = 0\, , \\
\label{eq:dS}
d_\a S^{\a\b\g}(x) &=& \partial_\a S^{\a\b\g}+ \Gamma^\a_{\a\lambda} S^{\lambda\b\g} 
 + \Gamma^\b_{\a\lambda} S^{\a\lambda\g} + \Gamma^\g_{\a\lambda} S^{\a\b\lambda} = 0,
\ea
respectively, where $\Gamma^\b_{\a\lambda}$ are Christoffel symbols~\cite{Singh:2020rht}, and the conformal transformation of the Christoffel symbols is~\cite{Faraoni:1998qx,Loganayagam:2008is,Singh:2020rht}
\beq
 \Gamma_{\lambda\a}^{\b}= \hat{\Gamma}_{\lambda\a}^{\b} + \delta^{\b}_{\lambda}\partial_{\a}\varphi+ \delta^{\b}_{\a}\partial_{\lambda}\varphi-
 \hat{g}_{\lambda\a}\hat{g}^{\b\sigma}\partial_{\sigma}\varphi \, ,
 \label{eq:dG}
\eeq
with $\delta^{\b}_{\lambda}$ denoting the Kronecker delta function and $\varphi$ being the function of spacetime coordinates.
For 4D spacetime, net baryon number~(\ref{eq:dN}) conservation equation is conformal-frame independent~\cite{Baier:2007ix,Bhattacharyya:2007vs,Loganayagam:2008is,Gubser:2010ze,Gubser:2010ui}, and since net baryon number has conformal weight $\Delta_{N^\a}=4$, one obtains
\beq
d_\a N^{\a} =\Omega^4  \hat{d}_{\a}\hat{N}^{\a}.
\label{eq:ConsN}
\eeq
Now putting \rf{eq:ConfN} and \rf{eq:dG} in \rf{eq:dT},
we obtain the transformation of conservation of energy and linear momentum~\cite{Bhattacharyya:2007vs,Loganayagam:2008is}
\beq
d_\a T^{\a\b} =\Omega^6 \left[\hat{d}_{\a}\hat{T}^{\a\b}- \hat{T}^{\lambda}_{\HP\lambda}\hat{g}^{\b\delta}\partial_\delta\varphi\right]\,.
\label{eq:ConsT}
\eeq 
One observe that $\hat{T}^{\a\b}$ should have trace 0 to be conserved in de Sitter spacetime~\cite{Callan:1970ze,DiFrancesco:1997nk,Forger:2003ut}, and using \rf{eq:ConfN} and \rf{eq:dG} in \rf{eq:dS}, spin conservation law transforms as
\beq
d_\a S^{\a\b\g} =\Omega^6 \left[\hat{d}_{\a}\hat{S}^{\a\b\g}(\hat{S}_{\lambda}^{\HP\lambda\g}\hat{g}^{\b\sigma} + \hat{S}^{\a\b}_{\HP\HP\a}\hat{g}^{\sigma\g})\partial_\sigma\varphi \right].\label{eq:ConsS}
\eeq
We find from above equation that spin tensor must satisfy the condition $\hat{S}_{\a}^{\HP\a\b}=0$ in order to have conformal invariance of spin conservation law.
It is easy to see that this condition is not satisfied by GLW (\ref{eq:GLW}) and HW \rfn{eq:HW} definitions, and hence, conformal invariance of \rf{eq:ConsS} breaks.
\section{Summary}
We have analysed the properties of the laws of conservation for net baryon number, energy-linear momentum, and spin with respect to the conformal transformations. We found the condition required for the conformal invariance of the spin conservation law, which the GLW and HW spin tensors explicitly breaks.

I am grateful to Radoslaw Ryblewski and Gabriel Sophys for their fruitful collaboration and thank W. Florkowski, D. S\'en\'echal and M. Shokri for inspiring discussions. Supported in part by the Polish National Science Center Grants No. 2016/23/B/ST2/00717 and No. 2018/30/E/ST2/00432.
%

\end{document}